\documentclass[12pt]{iopart}

\usepackage{graphicx}
\usepackage{dcolumn}% Align table columns on decimal point
\usepackage{bm}% bold math
\usepackage{amsthm,amssymb,ascmac}
\usepackage{cite}
\usepackage{color}

\usepackage[normalem]{ulem}

\usepackage[hang,small,bf]{caption}
\usepackage[subrefformat=parens]{subcaption}
\newcommand{\be}{\beta}

\newcommand{\bea}{\begin{eqnarray}}
\newcommand{\eea}{\end{eqnarray}}
\newcommand{\aeq}{&=&}

\newcommand{\aeqd}{&:= &}

\newcommand{\aeqle}{& \le &}
\newcommand{\aeqge}{& \ge &}
\newcommand{\bra}{\langle}
\newcommand{\dbra}{\langle  \hspace{-0.5 mm} \langle}
\newcommand{\ket}{\rangle}
\newcommand{\dket}{\rangle \hspace{-0.5 mm} \rangle }

\newcommand{\mL}{\mathcal{L}}

\newcommand{\mD}{\mathcal{D}}

\newcommand{\mA}{\mathcal{A}}
\newcommand{\mB}{\mathcal{B}}

\newcommand{\mW}{\mathcal{W}}
\newcommand{\dg}{^\dagger}
\newcommand{\tl}{\tilde}

\newcommand{\defe}{:=}
\newcommand{\ep}{\varepsilon}
\newcommand{\ga}{\gamma}
\newcommand{\Ga}{\Gamma}

\newcommand{\al}{\alpha}

\newcommand{\sig}{\sigma}
\newcommand{\f}{\frac}
\newcommand{\half}{\frac{1}{2}}
\newcommand{\pr}{\prime}

\newcommand{\Dl}{\Delta}

\newcommand{\om}{\omega}

\newcommand{\com}{\hspace{0.5mm}, \quad}
\newcommand{\la}{\label}
\newcommand{\ke}[1]{\vert #1 \rangle }
\newcommand{\abs}[1]{\vert #1 \vert }
\newcommand{\no}{\nonumber}

\newcommand{\re}[1]{(\ref{#1})}
\newcommand{\res}[1]{\S \ref{#1}}

\newcommand{\hs}{\hspace}

\newcommand{\RM}[1]{{\rm{#1}}}

\newcommand{\co}[1]{``{#1}''}

\newcommand{\brv}[1]{\langle #1 \vert}
\newcommand{\dbr}[1]{ \langle  \hspace{-0.5 mm} \langle #1 \vert }
\newcommand{\dke}[1]{\vert #1 \rangle \hspace{-0.5 mm} \rangle }
\newcommand{\nor}[1]{\vert\vert #1 \vert\vert}
\newcommand{\rd}[1]{\textcolor{black}{#1}}

\begin{document}

\title[Speed limits of the trace distance for open quantum system]{Speed limits of the trace distance for open quantum system}

\author{Satoshi Nakajima \& Yasuhiro Utsumi}

\address{Department of Physics Engineering, Faculty of Engineering, Mie University, 
Tsu, Mie 514-8507, Japan}
\ead{nakajima@eng.mie-u.ac.jp}
\vspace{10pt}
\begin{indented}
\item[]\today
\end{indented}

\begin{abstract}
We investigate the speed limit of the state transformation in open quantum systems described by the Lindblad type quantum master equation. 
We obtain universal bounds of the total entropy production described by the trace distance between the initial and final states in the interaction picture. 
Our bounds can be tighter than the bound of Vu and Hasegawa [Phys. Rev. Lett. \textbf{126}, 010601 (2021)] which measures the distance by the eigenvalues of the initial and final states: 
This distance is less than or equal to the trace distance.
For this reason, our results can significantly improve Vu-Hasegawa's bound. 
The trace distance in the Schr\"{o}dinger picture is bounded by a sum of the trace distance in the interaction picture and the trace distance for unitary dynamics described by only the Hamiltonian in the quantum master equation. 
\end{abstract}

% Uncomment for keywords
\vspace{2pc}
\noindent{\it Keywords}: speed limit, trace distance, open quantum system, quantum master equation
%
% Uncomment for Submitted to journal title message
%\submitto{\JPA}
%
% Uncomment if a separate title page is required
%\maketitle
% 
% For two-column output uncomment the next line and choose [10pt] rather than [12pt] in the \documentclass declaration
%\ioptwocol
%

\section{Introduction} 

In recent years, studies of time-dependent open systems have been active \cite{Bounds}. 
These studies relate to quantum pumps \cite{Nakajima15, Takahashi2020}, excess entropy production \cite{Sagawa, Nakajima17, Nakajima17D}, 
the information geometric approach \cite{Ito18, E7},  the efficiency and power of heat engines \cite{Tajima, Kamimura, Averin17}, 
shortcuts to adiabaticity \cite{RMP19, Takahashi17, E8, Funo2020, E9}, and speed limits 
\cite{E9, E1, E2, E3, review,  Shiraishi18, Funo19, Nakajima21, Vu21, Vu22, Vu22b, Takahashi2021, E6}. 
Obtaining a fundamental bound on the speed of state transformation is an important issue relevant to broad research fields including quantum control theory \cite{textbook1}
and foundations of nonequilibrium statistical mechanics \cite{textbook2}.
Speed limits for time-dependent closed quantum systems have been studied for more than a half-century \cite{Bounds}. 
Since 1945, the Mandelstam-Tamm relation \cite{Mandelstam} $\mL/\int_0^\tau dt \ \Dl E \le 1$ has been known 
(\ref{A_E}. In this paper, we set $\hbar=1$). 
Here, $\mL$ is the distance between the initial and final states \rd{(the Bures angle, see \ref{A_E})},
 and $\Dl E$ is the energy fluctuation. 

\rd{
About a decade ago, speed limits of open quantum systems had been intensively studied by adopting various distance measures between two quantum states \cite{E1, E2, E3, review}. 
Reference \cite{E1} derived the upper bound of the Bures angle expressed by the quantum Fisher information. 
For systems described by the quantum master equation $d\rho/dt = \hat{K}(\rho)$ ($\rho(t)$ is the density operator (the state) of the system), Ref.\cite{E2} provided the upper bound of the relative purity $\tr[\rho(t)\rho(0)]/\tr[\rho(0)^2]$  expressed with the adjoint of the generator of the dynamical map $\hat{K}$. 
Reference \cite{E3} estimated Margolus-Levitin-type and Mandelstam-Tamm-type bounds by using the Bures angle for pure initial state and several norms of $\hat{K}(\rho)$. 
A review of quantum speed limits for closed and open systems until around 2017 is given by Ref.\cite{review}. 
More recently, the speed limits on observables of open quantum systems are discussed~\cite{E6}.}

Recently, even in classical systems, it turns out that there exist speed limits expressed in terms of the distance between states \cite{Shiraishi18}. 
\rd{Remarkably, the classical speed limits connect the distance and the thermodynamic entropy.}
Shiraishi \textit{et al}. \cite{Shiraishi18} demonstrated that 
\bea
\sig \ge \f{l^2}{\int_0^\tau dt \ 2A_\RM{c}(t)} \la{Shiraishi_relation}
\eea
for a system described by a classical master equation 
$\f{d}{dt}p_n(t) =\sum_m W_{nm}p_m(t)$. 
$W_{nm}$ is the transition matrix satisfying the local detailed balance condition \cite{local_detailed_balance} and $p_n(t)$ is the probability of state $n$ at time $t$. 
$\sig$ is the total entropy production, 
the distance $l \defe \sum_n \abs{p_n(\tau)-p_n(0)}$ is the $L^1$ norm, 
and $A_\RM{c}(t)\defe \sum_{n \ne m}W_{nm}p_m(t)$ is the activity (total number of transitions per unit time).

The speed limits \rd{in terms of the entropy production} for the open quantum systems described by the Lindblad type quantum master equation 
(the Gorini-Kossakowski-Sudarshan-Lindblad equation), 
\bea
\f{d}{dt} \rho(t) \aeq -i[H(t),\rho(t)]+\mD(\rho(t)) , \la{QME}
\eea
have been researched actively \rd{in recent years. 
Here,} $H(t)\defe H_S(t)+H_L(t)$, where $H_S$ is the system Hamiltonian and $H_L$ is the Lamb shift Hamiltonian, 
which satisfies $[H_L(t), H_S(t)] = 0$.  
$\mD(\rho)$ represents dissipation and is given by 
$\mD(\rho)=\sum_k \ga_k \hat{D}[L_k](\rho)$ with 
$\hat{D}[X](Y) \defe \big(XYX\dg -\half X\dg X Y  -\half Y X\dg X\big)$.
In this paper, $X$ and $Y$ denote linear operators of the system.  
$\ga_k$ are non-negative real numbers which describe the strength of the dissipation.
The label $k$ is a tuple $(b,a,\om)$ where $b$ is the label  of the bath.
The jump operators $L_{b, a,\om}$ satisfy
\bea
[L_{b,a,\om},H_S] = \om L_{b,a,\om} \com L_{b,a,-\om}=L_{b,a,\om}\dg .\la{Lind1}
\eea
We assume the local detailed balance condition
\bea
\ga_{b,a,-\om} \aeq e^{-\be_b \om}\ga_{b,a,\om} , \la{Lind2}
\eea
where $\be_b$ is the inverse temperature of the bath $b$. 
Note that $L_k$, $\om$, $\ga_k$ and $\be_b$ can depend on time.
The total entropy production is given by $\sig \defe \int_0^\tau dt \ \dot{\sig}$ where 
\bea
\dot{\sig} \aeqd -\tr\Big[\f{d\rho}{dt}\ln \rho\Big]-\sum_b \be_b \tr[\mD_b(\rho) H_S]
\eea
is the entropy production rate. 
Here, $\mD_b(\rho) $ denotes the contribution from the bath $b$ of $\mD(\rho)$.

For the system described by \re{QME}, there are two approaches to speed limits. 
The first approach is Funo \textit{et al}.'s approach \cite{Funo19}, which treats the first and second terms of the right-hand side of \re{QME} equally. 
Funo \textit{et al}. \cite{Funo19} demonstrated  that
\bea
\nor{\rho(\tau)-\rho(0)}_1  \aeqle c_1+c_2+c_3 , \la{Funo_relation} 
\eea
with 
\bea
c_1\aeqle 2\int_0^\tau dt\ \Dl E, \la{c_1} \\
c_3\aeqle \sqrt{2\sig \int_0^\tau dt \ \mA(t)}. 
\eea
Here, $\nor{\rho(\tau)-\rho(0)}_1$ is the trace distance and $\nor{X}_1\defe \tr\sqrt{X\dg X}$ is the trace norm.  
$c_1$ corresponds to the contribution from the first term of the right-hand side of \re{QME}, $c_2$ and $c_3$ correspond to the contribution from the second term of the right-hand side of \re{QME} \cite{c_i}. 
$\Dl E \defe \sqrt{\tr(\rho(t) H(t)^2) -[\tr(\rho(t) H(t)) ]^2}$ is the energy fluctuation. 
$\mA(t)$ is defined by
\bea
\mA(t) \aeqd \sum_{n \ne m} \mW_{mn} p_n(t) \la{mA}
\eea
with $\mW_{mn} \defe  \sum_k\ga_k \abs{\brv{m(t)}L_k \ke{n(t)}}^2$. 
Here, we used the spectral decomposition of $\rho(t)$:
\bea
\rho(t)=\sum_n p_n(t)\ke{n(t)}\brv{n(t)} . \la{SD}
\eea
If the quantum master equation reduces to the classical master equation \cite{Q_to_C}, 
 \re{Funo_relation} reduces to \re{Shiraishi_relation} because $c_1=c_2=0$.
For no dissipation limit $\ga_k = 0$, \re{Funo_relation} becomes a Mandelstam-Tamm type relation because of $c_2=c_3=0$. 

The second approach is Vu's approach \cite{Vu21,Vu22}, which focuses on the second term of the right-hand side of \re{QME}. 
Vu and Hasegawa \cite{Vu21} demonstrated that
\bea
\sig \ge \sig_\RM{V1} \defe \f{d_T(\rho(\tau), \rho(0))^2}{\int_0^\tau dt \ 2B(t)} . \la{Vu-Hasegawa_relation}
\eea
Here,  
\bea
B(t)\aeqd \tr \Big[\rho(t) \sum_k \ga_k L_k\dg L_k \Big]  \no\\
\aeq \mA(t) + \sum_{n}\sum_k p_n(t)\ga_k \abs{\brv{n(t)}L_k \ke{n(t)}}^2
\eea
corresponds to the  activity \cite{comment1} 
\rd{(a similar quantity appears in Refs.\cite{E11a, E12} in the context of decoherence times)}. 
$d_T$ is defined by 
$d_T(\rho(\tau), \rho(0)) \defe \sum_n \abs{b_n -a_n}$ where $\{a_n \}$ and  $\{b_n \}$ are increasing eigenvalues of $\rho(0)$ and $\rho(\tau)$.
For no dissipation limit, \re{Vu-Hasegawa_relation} is consistent because $d_T(\rho(\tau), \rho(0))=0$ holds with  $B(t)=0$ and $\sig=0$. 
\re{Vu-Hasegawa_relation} is improved as \cite{Vu22b}
\bea
\sig \ge \sig_\RM{V0} \defe \f{d_T(\rho(\tau), \rho(0))^2}{\int_0^\tau dt \ 2M(t)} 
\eea
with 
\bea
M(t) \aeqd \sum_k \sum_{m \ne n} \Psi(a^{(k)}_{mn}, a^{(-k)}_{nm}) ,\\
a^{(k)}_{mn} \aeqd \ga_k \abs{\brv{m(t)}L_k \ke{n(t)}}^2 p_n(t).
\eea
Here, we used \re{SD} and $-k\defe (b,a,-\om)$. 
$\Psi(\al,\be)$ is the logarithmic mean of $\al$ and $\be$ given by
$\Psi(\al,\be) \defe (\be-\al)/(\ln \be/\al)$ $(\al \ne \be)$ and $\Psi(\al, \al) \defe \al$. 
The relation $\Psi(\al, \be)\le (\al+\be)/2$ leads to $M(t) \le \mA(t) \le B(t)$ and thus $\sig_\RM{V0} \ge \sig_\RM{V1}$. 

For a system of which Hilbert space is $d$-dimensional, 
Vu and Saito \cite{Vu22} demonstrated that 
\bea
\sig \ge \f{\nor{ \rho(\tau)-\rho_0}_1^2}{\int_0^\tau dt \ 2B(t)} \la{Vu-Saito_relation}
\eea
under the condition that the initial state is completely mixed as $\rho(0)=\rho_0 \defe 1/d$.
For no dissipation limit, \re{Vu-Saito_relation} is also consistent because of $\nor{ \rho(\tau)-\rho_0}_1=0$.

We consider \re{Vu-Hasegawa_relation} and \re{Vu-Saito_relation} possess the following shortcomings, which we would like to improve in the present paper. 
(i) $d_T$ can be zero between different states: 
When there is an unitary operator $U_\tau$ such that $\rho(\tau)=U_\tau \rho(0) U_\tau \dg $, $d_T$ becomes zero and thus cannot distinguish between the two states. 
(ii) Even in the classical master equation limit \cite{Q_to_C}, \re{Vu-Hasegawa_relation} does not lead to \re{Shiraishi_relation}: $d_T(\rho(\tau), \rho(0))$ does not become $l$
\cite{d_V} and $B(t)>A_\RM{c}(t)$ in general.
(iii) In \re{Vu-Saito_relation}, we can not replace $\rho_0$ by an any initial state $\rho(0)$.
In fact, in the weak dissipation limit $\ga_k \to 0$, although $\sig $ and $B(t)$ vanish, $\nor{ \rho(\tau)-\rho(0)}_1$ remains. 

The structure of the paper is as follows. 
First, we summarize our main results (\res{s_main}). 
Next, we explain derivations (\res{s_Derivation}). 
We apply our speed limits to a general system of which Hilbert space is two-dimensional (\res{s_3.1}): 
It includes a spinless quantum dot coupled to a single lead (\res{s_3.2}) and a qubit system (\res{s_Qubit}). 
In \res{s_Summary}, we summarize this paper. 
In \ref{A_E}, we derive the Mandelstam-Tamm relation for mixed state. 
\ref{App_B} is for the detailed calculations for \res{s_Derivation}. 
\rd{In \ref{s_A}, we derive a bound for the trace distance in the Schr\"{o}dinger picture. }
We prove $d_V \ge d_T$ in \ref{A_D}. 
\ref{A_Qubit} is for the detailed calculations for \res{s_Qubit}. 

\section{Main results} \la{s_main}

The main results of this paper are
\bea
\sig \aeqge \sig_0 \ge \sig_1 \ge  \sig_2 , \la{Goal} \\
\sig_0 \aeqd \f{\nor{\tl \rho(\tau)-\tl \rho(0)}_1^2}{\int_0^\tau dt \ 2A_\varphi(t)} ,\la{Goal_0}\\
\sig_1 \aeqd \f{\nor{\tl \rho(\tau)-\tl \rho(0)}_1^2}{\int_0^\tau dt \ [B(t)+B^\pr(t)]} ,\\
\sig_2 \aeqd \f{\nor{\tl \rho(\tau)-\tl \rho(0)}_1^2}{\int_0^\tau dt \ [B(t)+B_\infty(t)]},
\eea
where $\tl \rho(t)\defe U\dg(t)\rho(t)U(t)$ denotes the interaction picture. 
Here, $U(t)$ is defined by $\f{d}{dt}U(t)= -iH(t)U(t)$ and $U(0)=1$. 
$A_\varphi$, $B$, $B^\pr$, and $B_\infty$ are quantum extensions of the activity. 
$A_\varphi(t)$ is given by 
\bea
A_\varphi(t) \aeqd \tr\Big( \tl \rho(t) \f{1}{4}\sum_k \ga_k [ \varphi , \tl L_k]\dg [ \varphi , \tl L_k] \Big) .\la{A_varphi}
\eea
$ \varphi(t)$ is  defined by 
\bea
 \varphi(t) \defe\Phi(\tl \rho(t)-\tl \rho(0)) .
\eea
Here, $\Phi$ maps a self-adjoint operator $X$ to a self-adjoint operator as $\Phi(X) \defe \sum_n \RM{sign}(x_n) \ke{n}\brv{n}$, 
where the spectral decomposition of $X$ is $X=\sum_n x_n \ke{n}\brv{n}$. 
$\RM{sign}(x)$ is the sign of $x$. 
$B^\pr(t)$ and $B_\infty(t) $ are defined by
\bea
B^\pr(t) \aeqd \tr\Big(\varphi \tl \rho\varphi \sum_k \ga_k \tl L_k \dg \tl L_k\Big)  , \\
B_\infty(t) \aeqd \sum_k \ga_k \nor{L_k}_\infty^2 \ge B^\pr(t),
\eea
where $\nor{L_k}_\infty^2 $ equals to the maximum eigenvalues of $L_k \dg L_k$.
$\nor{Y}_\infty$ is called the spectral norm. 

The second inequality of \re{Goal} $\sig \ge \sig_1$ leads to \re{Vu-Saito_relation} for  $\rho(0)=\rho_0 $ because $B^\pr(t)=B(t)$ and  $\nor{\tl \rho(\tau)-\tl \rho_0}_1=\nor{ \rho(\tau)-\rho_0}_1$ hold
and thus $\sig_1=\sig_\RM{V1}$.
In the classical master equation limit, the first inequality of \re{Goal} $\sig \ge \sig_0$ leads to \re{Shiraishi_relation}: In this limit, $\nor{\tl \rho(\tau)-\tl \rho(0)}_1=l$ and $A_\varphi \le A_c$ hold (\res{A_B}) 
and thus $\sig_0$ is larger than the right-hand side of \re{Shiraishi_relation}.
Even for no dissipation limit, \re{Goal} is consistent because $\nor{\tl \rho(\tau)-\tl \rho(0)}_1=0$ holds with $\sig=0$ and $A_\varphi(t)=B(t)=B^\pr(t)=B_\infty(t)=0$.

We notice that
\bea
d_T(\tl \rho(\tau), \tl \rho(0)) \le \nor{\tl \rho(\tau)-\tl \rho(0)}_1 \la{d_T_2022}
\eea
for finite dimensional Hilbert space (p.512 in Ref.\cite{NC})
and $d_T(\tl \rho(\tau),\tl \rho(0)) =d_T(\rho(\tau),\rho(0)) $. 
\re{d_T_2022} can be also derived from \cite{Vu22b}
\bea
d_T(\rho_2, \rho_1) \aeq \min_{V \in \{U\vert U\dg U=1\}} \nor{V\rho_2V\dg-\rho_1}_1.
\eea
\re{d_T_2022} indicates that our bounds $\sig_k$ $(k=0,1,2)$ can be better than \re{Vu-Hasegawa_relation}. 

Similarly to \re{Funo_relation}, we can separate contributions from unitary dynamics and dissipation. 
By the triangle inequality, the trace distance in the Schr\"{o}dinger picture is bounded as \cite{comment_dis_s}:
\bea
\nor{\rho(\tau)-\rho(0)}_1 \aeqle \nor{\check{\rho}(\tau)- \rho(0)}_1+\nor{\rho(\tau)-\check{\rho}(\tau)}_1 \no\\
\aeq \nor{\check{\rho}(\tau)- \rho(0)}_1+\nor{\tl \rho(\tau)-\tl \rho(0)}_1. \la{tr_distance_s}
\eea
Here, $\check{\rho}(t) \defe U(t)\rho(0) U\dg(t)$. 
The first term of the right-hand side of \re{tr_distance_s} is related to the unitary time evolution and is bounded by the Mandelstam-Tamm type relation
 (\ref{A_E}, \cite{Funo19})
\bea
\nor{\check{\rho}(\tau)- \rho(0)}_1 \le 2\int_0^\tau dt \ \Dl \check{E}.
\eea
Here, $\Dl \check{E} \defe \sqrt{\tr(\check{\rho}(t) H(t)^2) -[\tr(\check{\rho}(t) H(t)) ]^2}$.
Using \re{Goal} and \re{Goal_0}, the second term of the right-hand side of \re{tr_distance_s} is bounded as
\bea
\nor{\tl \rho(\tau)-\tl \rho(0)}_1 \le \sqrt{2\sig \int_0^\tau dt \ A_\varphi(t)}.
\eea
\rd{In \ref{s_A}, we explain an alternative bound for the trace distance in the Schr\"{o}dinger picture
which does not refer to the virtual isolated system. }

We discuss the meaning of the two distances when the Hilbert space is two-dimensional.   
In this case, the state of the system can be written as $\tl \rho(t)=\half(1+\bm{r}(t)\cdot \bm{\tau})$.
Here, $\bm{\tau}=(\tau_x,\tau_y,\tau_z)$, $\tau_i$ is the Pauli matrix, and  $\bm{r}(t)$ is the Bloch vector. 
The trace distance and $d_T$ are given by
\bea
\nor{\tl \rho(\tau)-\tl \rho(0)}_1\aeq \abs{\bm{r}(\tau)-\bm{r} (0)} ,\la{trace distance_B}\\
d_T(\tl \rho(\tau), \tl \rho(0)) \aeq \abs{\abs{\bm{r}(\tau)}-\abs{\bm{r}(0)} } \la{d_T_B},
\eea
with $\abs{\bm{x}} \defe \sqrt{\bm{x} \cdot \bm{x}}$. 
$d_T$ measures the difference between the length of the two Bloch vectors and 
does not quantify the coherence.

\section{Derivation of the main results} \la{s_Derivation}

Our key idea is the use of the trace distance in the interaction picture within Vu's framework \cite{Vu21, Vu22}. 
We introduce a semi-inner product by
\bea
\dbra X, Y\dket_{\tl \rho, k} \aeqd \tr \big[X\dg \{\tl \rho \}_k(Y) \big] \la{semi_def}
\eea
with 
\bea
\{\rho \}_k(X) \aeqd \int_0^1 ds \  (\ga_{-k}\rho)^s X (\ga_k \rho)^{1-s}.
\eea
Here, $\ga_{-k}= \ga_{b,a,-\om}$. 
$\{\rho \}_k(X)$ is the logarithmic mean transformation \cite{Matrix_analysis_book}.
The semi-inner product satisfies $\big(\dbra X, Y\dket_{\tl \rho, k} \big)^\ast = \dbra Y, X\dket_{\tl \rho, k} $ and $\nor{X}_{\tl \rho, k}^2 \defe \dbra X, X\dket_{\tl \rho, k} \ge 0$. 
This semi-inner product differs from Vu's semi-inner product $\tr[X\dg \hat{O}^{(b)}(Y)]$ where
$\hat{O}^{(b)}(X) \defe \half \sum_{a,\om}[\tl L_k\dg,\{\tl \rho \}_k([\tl L_k,X]) ]$ \cite{Vu21}.
The super-operator $\hat{O}^{(b)}$ corresponds to the Laplacian of a weighted graph \cite{Ito2022} 
(see \re{S25} and \res{A_entropy}). 
We consider the semi-inner product  \re{semi_def} provides more transparent descriptions (\ref{App_B}). 

Using the semi-inner product, we can obtain
\bea
\f{d}{dt} \nor{\tl \rho(t)-\tl \rho(0)}_1 \aeq \tr\Big[\varphi(t) \f{d\tl \rho}{dt} \Big] \no\\
\aeq \half \sum_k  \dbra [\tl L_k,\varphi(t)], [\tl L_k, -\ln \tl \rho-\be_b \tl H_S] \dket_{\tl \rho, k} .\la{key_1}
\eea
We used 
$\f{d}{dt}\tr[f(X(t))] =\tr\Big[f^\pr(X(t))\f{dX(t)}{dt}\Big]$ for a self-adjoint operator $X(t)$ and a differentiable function $f(x)$ in the first line.
In the second line, we used the quantum master equation in the interaction picture (\res{A_entropy})
\bea
\f{d}{dt}\tl \rho(t) \aeq \sum_k \ga_k \hat{D}[\tl L_k](\tl \rho) \no\\
\aeq \half \sum_k  [\tl  L_k\dg, \{\tl \rho\}_k([\tl L_k ,-\ln \tl \rho-\be_b \tl H_S] )] \la{S25}.
\eea
Further, \re{key_1} leads to 
\bea
\nor{\tl \rho(\tau)-\tl \rho(0)}_1 
\aeq \half \sum_k \int_0^\tau dt \  \dbra [\tl L_k,\varphi(t)], [\tl L_k, -\ln \tl \rho-\be_b \tl H_S] \dket_{\tl \rho, k}  \no\\
\aeqle \half \sum_k \int_0^\tau dt \ \sqrt{\Big\vert \Big\vert [\tl L_k,\varphi] \Big\vert\Big\vert_{\tl \rho, k}^2   \Big\vert \Big\vert [\tl L_k,-\ln \tl \rho-\be_b \tl H_S] \Big\vert\Big\vert_{\tl \rho, k}^2 }\no\\
\aeqle  \sqrt{\int_0^\tau dt \  \half \sum_k  \Big\vert \Big\vert [\tl L_k,\varphi] \Big\vert\Big\vert_{\tl \rho, k}^2} \cdot \sqrt{\sig} .\la{key_2}
\eea
Here, we used the Cauchy-Schwarz inequalities
\bea
\left \vert \dbra X, Y \dket_{\tl \rho, k} \right \vert \le \sqrt{\nor{X}_{\tl \rho, k}^2 \nor{Y}_{\tl \rho, k}^2 } 
\eea
and 
\bea
\sum_k \int_0^\tau dt \ \sqrt{\al_k(t)\be_k(t)} \le \sqrt{\sum_k \int_0^\tau dt \ \al_k(t)} \sqrt{\sum_k \int_0^\tau dt \ \be_k(t)} .
\eea
Here, $\al_k(t) $ and $\be_k(t)$ are non-negative real numbers. 
The entropy production rate can be written by using the semi-inner product \cite{Vu21}:
\bea
\dot{\sig}\aeq \half \sum_k \Big\vert \Big\vert [\tl L_k,-\ln \tl \rho-\be_b \tl H_S] \Big\vert\Big\vert_{\tl \rho, k}^2 .\la{sig_b}
\eea
In \re{key_2}, we can demonstrate  (\res{A_B})
\bea
\half \sum_k  \Big\vert \Big\vert [\tl L_k,\varphi] \Big\vert\Big\vert_{\tl \rho, k}^2 \le 2A_\varphi(t)  ,\la{B_B}
\eea
which leads to the tightest inequality of \re{Goal}.
We can demonstrate 
\bea
2A_\varphi(t) \le B(t)+B^\pr(t) \le B(t)+B_\infty(t) \la{eq41}
\eea
using $\tr\big( \tl \rho(t) \{\varphi , \tl L_k\}\dg \{\varphi , \tl L_k\} \big) \ge 0$, $\varphi(t)^2=1$, and $B_\infty(t) \ge B^\pr(t)$. 
Here, $\{X, Y\} = XY+YX$. 
Then, we obtain the other inequalities of \re{Goal}.

We compare the derivations of \re{Vu-Hasegawa_relation} and \re{Goal}. 
\re{Vu-Hasegawa_relation} can be derived as follows \cite{Vu21, Vu22}. 
For the spectral decomposition $\tl \rho(t)=\sum_n p_n(t)\ke{\tl n(t)}\brv{\tl n(t)}$, we put
\bea
\tl \phi(t) \defe \sum_n c_n(t)\ke{\tl n(t)}\brv{\tl n(t)} \com c_n(t)^2=1. \la{phi}
\eea
Then, 
$[\tl \rho(t), \tl \phi(t)]=0$ and  $\tl \phi(t)^2=1$ hold. 
For $c_n(t)=\RM{sign}(p_n(t)-p_n(0))$ or $c_n(t)=\RM{sign}(p_n(\tau)-p_n(0))$, 
\bea
d_V(\tl \rho(\tau),\tl \rho(0))\defe \sum_n \abs{p_n(\tau)-p_n(0)} 
= \tr \int_0^\tau dt \ \tl \phi(t) \f{d}{dt}\tl \rho(t) \la{d_T_int}
\eea
holds \cite{comment}. 
By repeating similar calculations from \re{key_2} and by exploiting $d_V(\tl \rho(\tau),\tl \rho(0)) \ge d_T(\tl \rho(\tau),\tl \rho(0))$ (\ref{A_D}) 
and $d_V(\tl \rho(\tau),\tl \rho(0)) =d_V(\rho(\tau),\rho(0))$,
we derive \re{Vu-Hasegawa_relation} (\res{A_B} and \res{A_C}). 
Note that $\tl \phi(t) \ne \varphi$ for any $c_n(t)$ in general.

\section{Application} \la{s_Application}

\subsection{General two-dimensional system} \la{s_3.1}

In this subsection, we consider a general system of which Hilbert space is two-dimensional. 
In general, the jump operators are written as \cite{jump_c}
\bea
\tl L_k \aeq (\bm{R}_k+i\bm{I}_k)\cdot \bm{\tau} .
\eea
Here, the components of $\bm{R}_k$ and $\bm{I}_k$ are real numbers. 
The equation of the motion of the Bloch vector of $\tl \rho$ is given by
\bea
\f{d\bm{r}}{dt} \aeq 2\sum_k \ga_k[-(\bm{R}_k^2+\bm{I}_k^2)\bm{r}+(\bm{R}_k\cdot \bm{r}) \bm{R}_k 
+(\bm{I}_k\cdot \bm{r} ) \bm{I}_k+2\bm{R}_k \times \bm{I}_k]. \la{eq_Bloch}
\eea
The activities are given by
\bea
A_\varphi \aeq \sum_k \ga_k\Big[\bm{R}_k^2-(\bm{\varphi} \cdot \bm{R}_k)^2
+\bm{I}_k^2-(\bm{\varphi} \cdot \bm{I}_k)^2 \la{A_p_2}
-2[\bm{\varphi} \cdot(\bm{R}_k \times \bm{I}_k)  ]\bm{\varphi} \cdot \bm{r} \Big] ,\\
B \aeq \sum_k \ga_k\Big[\bm{R}_k^2+\bm{I}_k^2-2(\bm{R}_k \times \bm{I}_k) \cdot \bm{r} \Big] ,\\
B^\pr \aeq \sum_k \ga_k\Big[\bm{R}_k^2+\bm{I}_k^2-2(\bm{R}_k \times \bm{I}_k) \cdot \bm{r}^\pr \Big] ,\\
B_\infty \aeq \sum_k \ga_k \Big[\bm{R}_k^2+\bm{I}_k^2+2\abs{\bm{R}_k \times \bm{I}_k}\Big].
\eea
Here, we expanded $\varphi$ as $\varphi=\bm{\varphi}\cdot \bm{\tau}$ with 
$\bm{\varphi}\defe \f{1}{\abs{\bm{r}-\bm{r}(0)}}[\bm{r}-\bm{r}(0)]$.
$\bm{r}^\pr=(x^\pr(t), y^\pr(t), z^\pr(t))$ is the Bloch vector of $\varphi(t) \tl \rho\varphi(t) $: 
\bea
\bm{r}^\pr =-\bm{r}+\f{2([\bm{r}-\bm{r}(0)]\cdot \bm{r})[\bm{r}-\bm{r} (0)]}{\abs{\bm{r}-\bm{r}(0)}^2}.
\eea
\re{A_p_2} is simplified as 
\bea
A_\varphi \aeq \f{B+B^\pr}{2}- \sum_k \ga_k\Big[(\bm{\varphi} \cdot \bm{R}_k)^2+ (\bm{\varphi} \cdot \bm{I}_k)^2 \Big].
\eea
Because $\bm{\varphi}$ depends on the initial state, $A_\varphi$ and $B^\pr$ depend on it. 
One can check for $\rho(0)=\rho_0$, \textit{i.e}., $\bm{r}(0)=0$, $B=B^\pr$, and $\nor{\tl \rho(\tau)-\tl \rho(0)}_1=d_T(\rho(\tau), \rho(0))$, and thus
$\sig_1=\sig_\RM{V1}$.

\subsection{Quantum dot} \la{s_3.2}

We analyze our inequality \re{Goal} for a spinless quantum dot coupled to a single lead \cite{Nakajima15, Nakajima17D}.
The quantum mater equation is given by
\bea
\f{d\rho}{dt} = -i[H_S, \rho] +\ga[1-f(\ep)]\hat{D}[a]( \rho)
+ \ga f(\ep)\hat{D}[a\dg](\rho) \la{QME_QD}
\eea
with $H_S=\ep a\dg a$. 
Here, $a$ is the annihilation operator of the electron of the system, 
$\ep$ is the energy level of  the system, 
$f(\ep) = \f{1}{e^{\be\ep}+1}$ is the Fermi distribution, 
$\be$ is the inverse temperature of the lead, and  
$\ga$ is the coupling strength. 
From \re{QME_QD}, we otain $L_1=2a$, $\ga_1=\f{1}{4}\ga[1-f(\ep)]$, $L_2=2a\dg$, and $\ga_2=\f{1}{4}\ga f(\ep)$. 
These lead to $\bm{R}_1=\bm{R}_2=(0,0,1)$, and $\bm{I}_1=-\bm{I}_2=(0,1,0)$. 
The equation of the motion of the Bloch vector $\bm{r}=(x,y,z)$ is given by
\bea
\f{d}{dt}x\aeq -\half \ga x \com
\f{d}{dt}y = -\half \ga y \com
\f{d}{dt}z=-\ga (z-[1-2f(\ep)]).
\eea
We calculate the trace distance and $d_T$ by \re{trace distance_B} and \re{d_T_B}. 
The activity $B(t)$ and its upper limit $B_\infty(t)$ are given by using $B(t)=\ga\big(1+[2f(\ep)-1]z(t)\big)/2$ and $B_\infty(t)=\ga$. 
$B^\pr(t)$ is obtained from $B(t)$ by replacing $z(t)$ with $z^\pr(t)$. 
$A_\varphi(t)$ is given by
\bea
A_\varphi(t) \aeq \f{\ga}{4}\Big(1+\f{[z-z(0)]^2}{\abs{\bm{r}-\bm{r}(0)}^2}+2[2f(\ep)-1]\f{[z-z(0)][\bm{r}-\bm{r}(0)]\cdot \bm{r} }{\abs{\bm{r}-\bm{r}(0)}^2}  \Big).
\eea
The entropy production is calculated as
\bea
\sig \aeq H_2\Big(\f{1+\abs{\bm{r}(\tau)}}{2} \Big)-H_2\Big(\f{1+\abs{\bm{r}(0)}}{2} \Big)
+\int_0^\tau dt \ \be J(t). \la{sig_2}
\eea
Here, $H_2(p) \defe -p \ln p - (1-p)\ln (1-p)$ is the binary entropy.  
The heat current $J(t)\defe -\tr[H_S\mD(\rho)]$ is given by
\bea
J(t) \aeq \half \ga\ep \Big([1-2f(\ep)]-z(t) \Big).
\eea
For $\ga t \gg 1$ and $\be \ep(t) \gg 1$ region, $B(t) \approx 0$ holds, however, $A_\varphi(t)$ and $B^\pr(t)$ remain finite in general.

\begin{figure}[t]
\begin{center}
  \begin{minipage}[b]{0.45\linewidth}
    \centering
    \includegraphics[keepaspectratio, width = 1 \columnwidth]{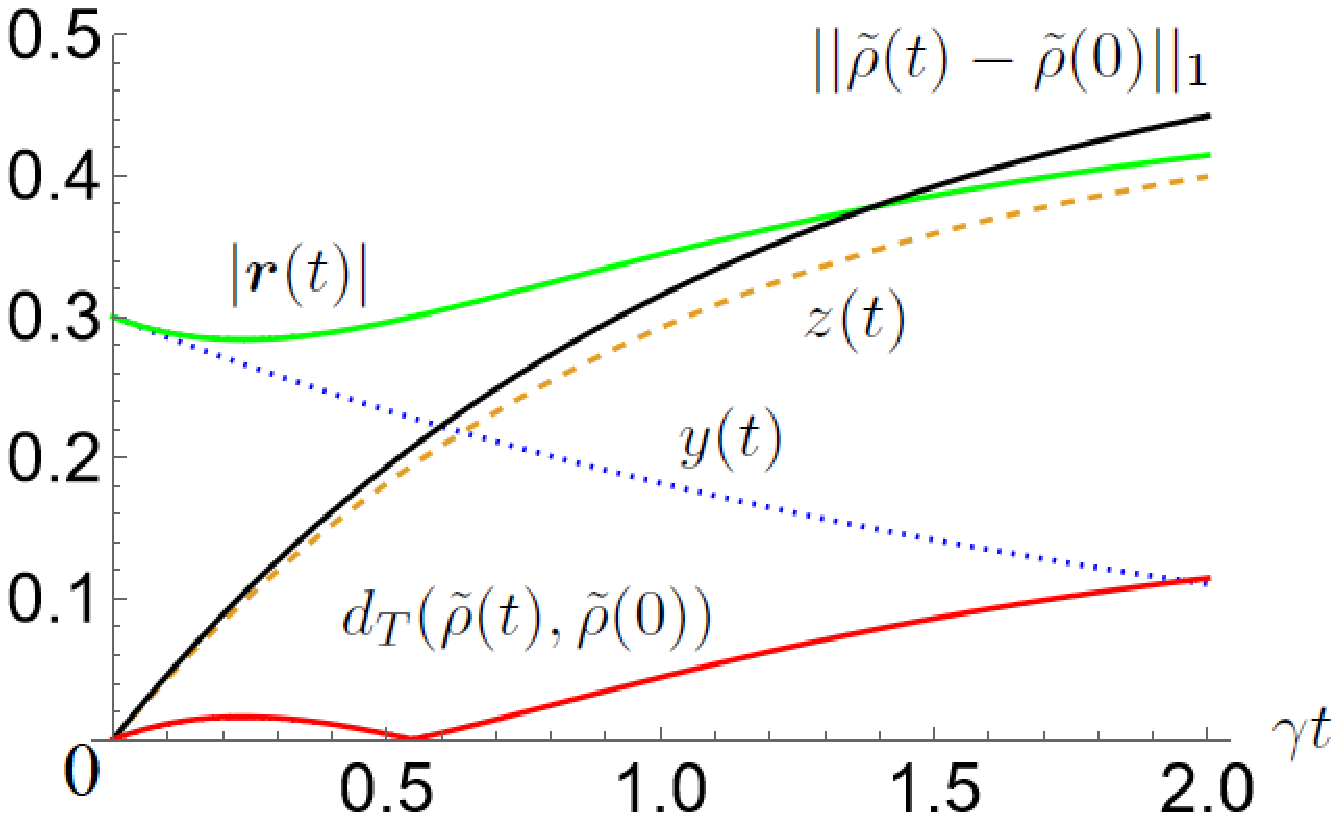}
    \subcaption{}
  \end{minipage}
  \begin{minipage}[b]{0.45\linewidth}
    \centering
    \includegraphics[keepaspectratio, width = 1 \columnwidth]{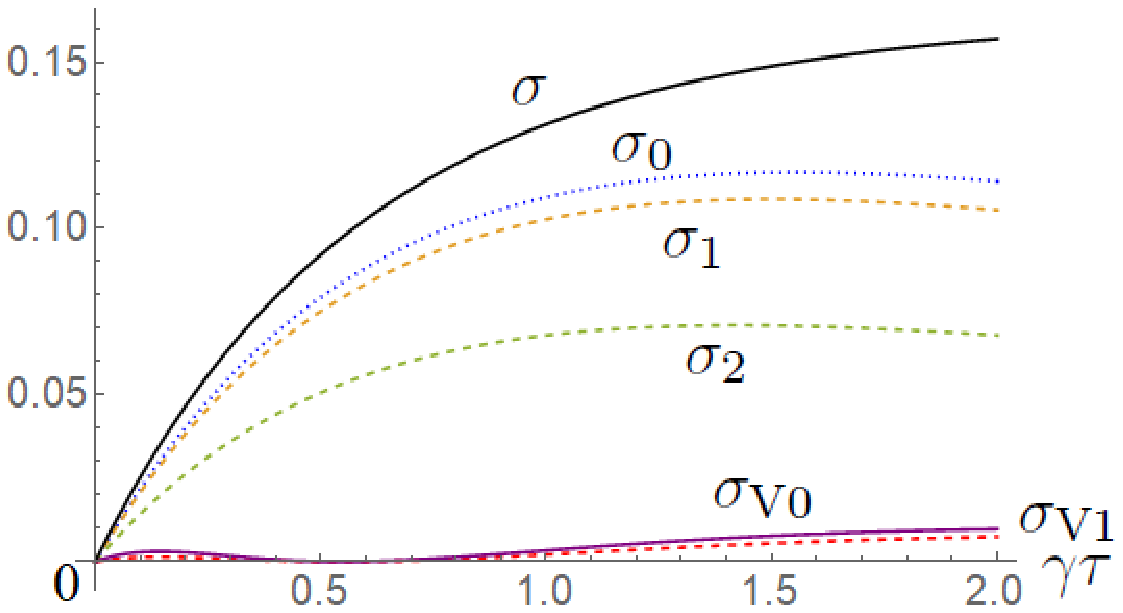}
    \subcaption{}
  \end{minipage}
  \caption{\label{Fig_1}(a) $\nor{\tl \rho(t)-\tl \rho(0)}_1$, $d_T(\tl \rho(t),\tl \rho(0))$, $y(t)$, $z(t)$, and $\abs{\bm{r}(t)}$. 
(b) The total entropy production $\sig$ and its bounds $\sig_0$, $\sig_1$, $\sig_2$, $\sig_\RM{V0}$, and $\sig_\RM{V1}$. 
The horizontal axes are time. 
$x(t)$ equals zero identically. 
$\be \ep=1$ and $\rho(0)=\half(1+0.3 \tau_y)$. 
}
\end{center}
\end{figure}

Figure \ref{Fig_1}(a) shows that the direction of the Bloch vector changes from the $y$-direction to the $z$-direction. 
In this process, the norm of the Bloch vector and then $d_T(\tl \rho(t), \tl \rho(0))$ change only slightly, see \re{d_T_B}.
At  $\ga t =0.547\cdots$,  although $\tl \rho(t) \ne \tl \rho(0)$, the distance $d_T(\tl \rho(t), \tl \rho(0))$ becomes zero. 
On the other hand, $\nor{\tl \rho(t)-\tl \rho(0)}_1^2$ is much larger than $d_T(\tl \rho(t),\tl \rho(0))^2$. 
Figure \ref{Fig_1}(b) shows that our bounds $\sig_0$, $\sig_1$ and $\sig_2$ are superior to Vu-Hasegawa's bound $\sig_\RM{V1}$. 
The Vu-Saito relation \re{Vu-Saito_relation}  is not applicable in this case.

\subsection{Qubit} \la{s_Qubit} 

We compare our bounds and Vu-Hasegawa's bound in the system studied in Ref.\cite{Vu22}.
We consider the qubit system described by the quantum master equation (see \res{A_Qubit} for detailed calculation)
\bea
\f{d\rho}{dt} \aeq -i[H_S(t),\rho]+\al \ga\ep(t) n(\ep(t))\hat{D}[\tau_+(\theta(t))](\rho) \no\\
&& +  \al \ga\ep(t) [ n(\ep(t))+1]\hat{D}[\tau_-(\theta(t))](\rho) \la{QME_Qubit}
\eea
with 
\bea
H_S(t)= \half \ep(t) \tau_z(\theta(t)).
\eea
$n(\ep) = \f{1}{e^{\be \ep} -1}$ is the Bose distribution. 
Here, 
\bea
\tau_\pm(\theta) \defe \half(\tau_x(\theta)\pm i\tau_y(\theta))
\eea
with 
\bea
\tau_x(\theta) \defe \tau_x \cos \theta -\tau_z \sin \theta , \ \tau_y(\theta) \defe \tau_y ,\  
\tau_z(\theta) \defe \tau_z \cos \theta +\tau_x \sin \theta.
\eea

\begin{figure}[t]
\begin{center}
  \begin{minipage}[b]{0.45\linewidth}
    \centering
    \includegraphics[keepaspectratio, width = 1 \columnwidth]{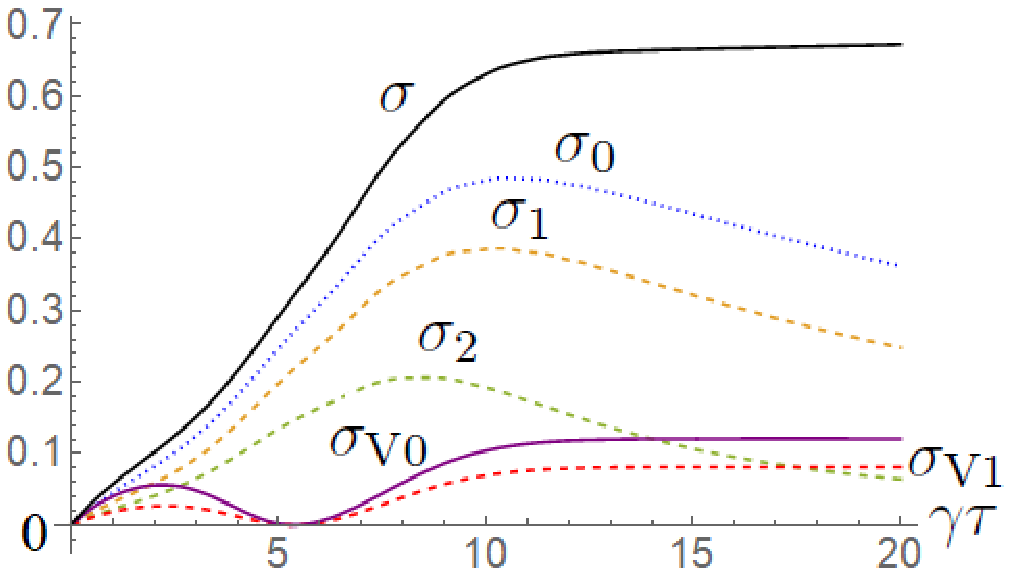}
    \subcaption{$\rho(0)=\half(1-0.5 \tau_y)$.}
  \end{minipage} 
  \begin{minipage}[b]{0.50\linewidth}
    \centering
    \includegraphics[keepaspectratio, width = 1 \columnwidth]{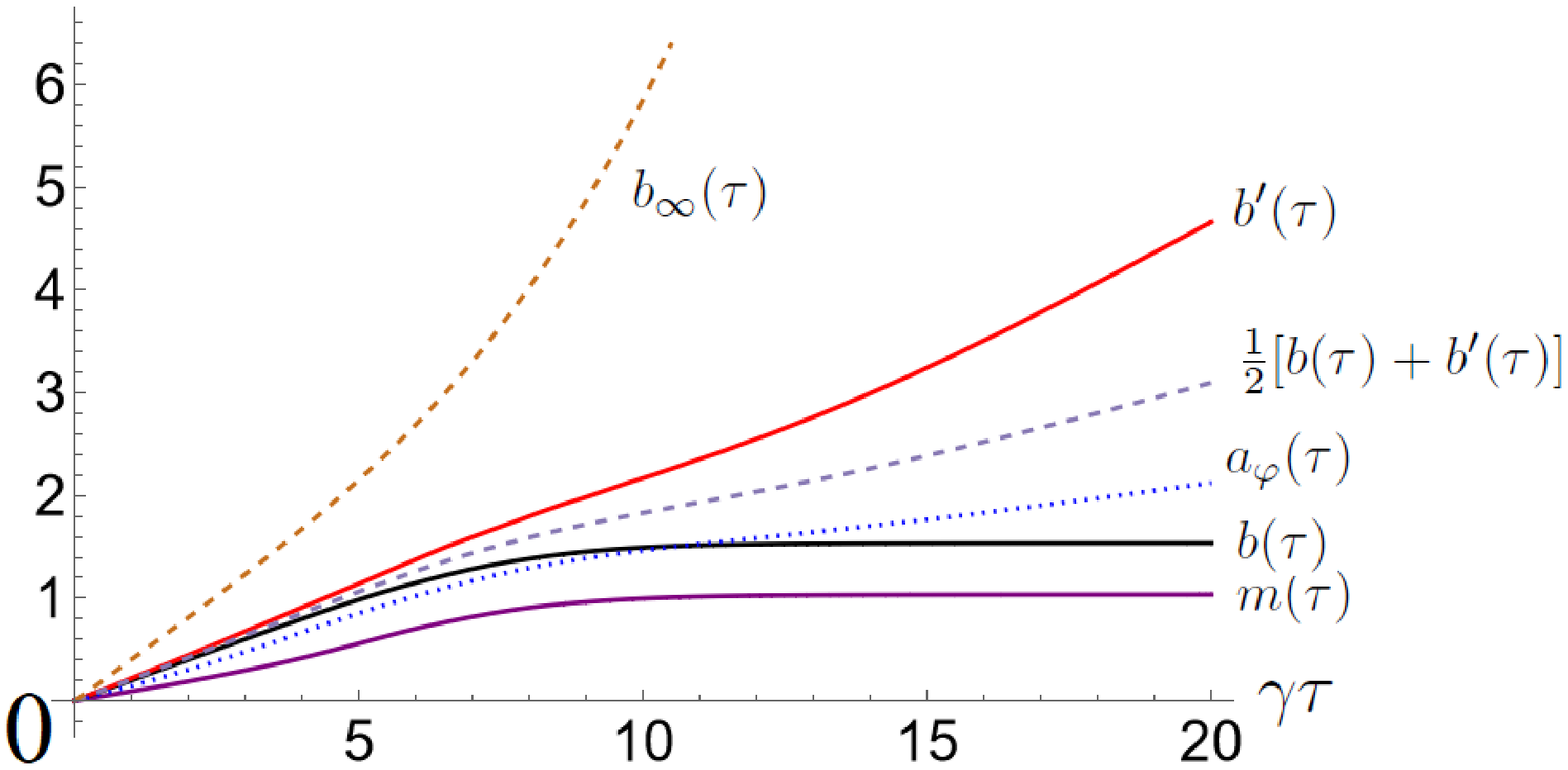}
    \subcaption{Activities for $\rho(0)=\half(1-0.5 \tau_y)$.}
  \end{minipage}
  
  \begin{minipage}[b]{0.45\linewidth}
    \centering
    \includegraphics[keepaspectratio, width = 1 \columnwidth]{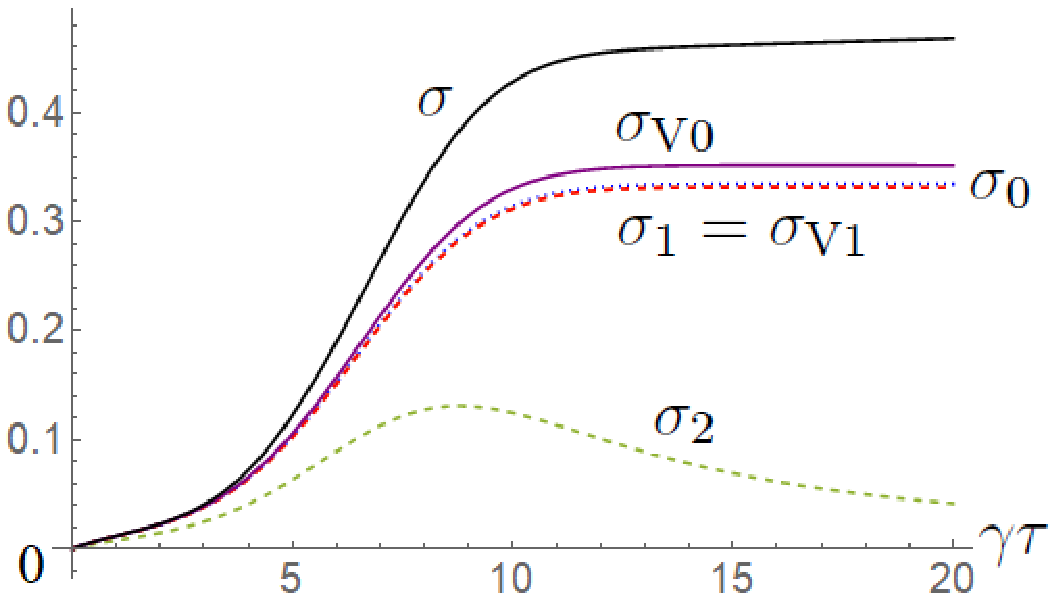}
    \subcaption{$\rho(0)=\half $.}
  \end{minipage}
  
\caption{\label{Fig_2}(a) $\sig$ and its bounds for $\rho(0)=\half(1-0.5 \tau_y)$. (b) $m(\tau)\defe \int_0^\tau dt \ M(t)$ ,
$b(\tau) \defe \int_0^\tau dt \ B(t)$,
$a_\varphi(\tau) \defe \int_0^\tau dt \ A_\varphi(t)$,
$b^\pr(\tau)\defe \int_0^\tau dt \ B^\pr(t)$,
$b_\infty(\tau)\defe \int_0^\tau dt \ B_\infty(t)$, and $\half[b(\tau)+b^\pr(\tau)]$ for $\rho(0)=\half(1-0.5 \tau_y)$. 
 (c) $\sig$ and its bounds for $\rho(0)=\half $. 
The horizontal axes are time. 
$\al=0.2 \be$, $\be \ga=1$, $\be \ep_0=0.4$, $\be \ep_1=10$, $\ga T=20$. 
In (a), for $\ga \tau< 0.15$, $\sig_\RM{V0}$ is larger than $\sig_0$. 
In (c), $\sig_0$ is slightly larger than $\sig_1$. 
}
\end{center}
\end{figure}

We consider the \co{nonoptimal protocol} in Ref.\cite{Vu22}: 
\bea
\theta(t) = \pi \Big(\f{t}{T}-1\Big) \com
\ep(t)= \ep_0+(\ep_1-\ep_0)\sin^2 \f{\pi t}{T}.
\eea
Figure \ref{Fig_2} shows $\sig$ and its bounds $\sig_0$, $\sig_1$, $\sig_2$, $\sig_\RM{V0}$, and $\sig_\RM{V1}$ 
for (a) $\rho(0)=\half(1-0.5 \tau_y)$ and (c) $\rho(0)=\half $ and (b) the integrals of the activities for $\rho(0)=\half(1-0.5 \tau_y)$. 
Figure \ref{Fig_2}(a) shows that our bounds $\sig_0$ and $\sig_1$ are superior  in a wide range. 
In Fig.\ref{Fig_2}(b), for $\ga t > 12$, $M(t)\approx 0$ and $B(t) \approx 0$ hold, however, 
$A_\varphi(t)$ and $B^\pr(t)$ remain finite because they depend on the initial state. 
If we fix $\ep(t)$ at $\ep_1$ after $t>T$ while $\theta(t)=\pi \big(\f{t}{T}-1\big)$, $\sig_0$ cross to $\sig_\RM{V1}$ at $\ga \tau \approx 120$. 
In this protocol, if $\bm{r}(0) \ne 0$ and $\be \ep_1 \gg 1$, our bounds get worse over time in general. 
For small $\ep(t)$, whether our bounds $\sig_0$ and $\sig_1$ or Vu-Hasegawa's bound $\sig_\RM{V1}$ decreases faster 
depends on the initial condition $\bm{r}(0)$. 
Figure \ref{Fig_2}(c) corresponds to the Fig. 1 (c) in Ref.\cite{Vu22}. 
In this case, $\sig_\RM{V0}$ is superior to $\sig_0$. 
In Fig.\ref{Fig_2}(c), not only $M(t)$ and $B(t)$ but also $A_\varphi(t)$ and $B^\pr(t) $ almost vanish for $\ga t > 12$.

\section{Summary} \la{s_Summary}

In open quantum systems described by the Lindblad type quantum master equation, 
we obtained universal bounds of the total entropy production described by the trace distance between the initial and final states. 
We considered the trace distance in the interaction picture instead of the distance of Ref.\cite{Vu21} (measured by the eigenvalues of the initial and final states) and trace distance in the Schr\"{o}dinger picture \cite{Vu22}. 
Our bounds can be tighter than the bound of Vu and Hasegawa \cite{Vu21}. 
Our results are applicable to an arbitrary initial state, beyound Vu-Saito's bound \cite{Vu22} applicable only to the completely mixed initial state. 
In the classical master equation limit, our tightest inequality leads to the inequality by Shiraishi \textit{et al}. \cite{Shiraishi18}. 
The trace distance in the Schr\"{o}dinger picture is bounded by a sum of the trace distance for unitary dynamics described by the system Hamiltonian and the trace distance in the interaction picture.

\subsection*{Acknowledgments}

We acknowledge helpful discussions with Kazutaka Takahashi. 
This work was supported by JSPSKAKENHI Grants No. 18KK0385, and No. 20H01827. 

\appendix

\section{Mandelstam-Tamm relation} \la{A_E}

In this section, we demonstrate that the Mandelstam-Tamm relation for mixed state \cite{E2}
\bea
\mL(\rho(\tau),\rho(0)) \le \int_0^\tau dt \ \Dl E \la{MT}
\eea
and 
\bea
\half \nor{\rho(\tau)-\rho(0)}_1 \le \int_0^\tau dt \ \Dl E \la{MT_type}
\eea
for a closed quantum system $S$ (described by $\f{d\rho}{dt} = -i[H(t),\rho]$). 
Here, $\mL(\rho,\sig) \defe \cos^{-1} F(\rho,\sig)$ is the Bures angle, 
$F(\rho,\sig)\defe \tr \sqrt{\sqrt{\rho} \sig \sqrt{\rho}}$ is the fidelity \cite{NC}, and
$\Dl E \defe \sqrt{\tr(\rho H^2)-[\tr(\rho H)]^2} $. 
First, we demonstrate \re{MT}. 
For  the spectral decomposition 
\bea
\rho(0) \aeq \sum_i p_i \ke{\phi_i}_S\brv{\phi_i},
\eea
we define a state
\bea
\dke{\rho(t)} \aeqd U(t) \sum_i \sqrt{p_i} \ke{\phi_i}_S \otimes \ke{i}_A .
\eea
Here, $U(t)$ is given by $dU(t)/dt=-iH(t)U(t)$ and $U(0)=1$. 
$\{\ke{i}_A\}$ is an orthonormal basis of the ancilla system $A$. 
Because $ \tr_A (\dke{\rho(t)} \dbr{\rho(t)}) =\rho(t)$ holds, 
$\dke{\rho(t)}$ is a purification of  $\rho(t)$. 
We consider the Robertson inequality
\bea
\Dl X \Dl E \ge \half \abs{\bra i[H, X] \ket_t} \la{RI}
\eea
with $X=\dke{\rho(0)}\dbr{\rho(0)} $. 
Here, $\Dl Y \defe \sqrt{\bra Y^2 \ket_t-\bra Y \ket_t^2} $ and $\bra Y \ket_t \defe \dbr{\rho(t)} Y\dke{\rho(t)} $. 
For a operator $Y_S$ of system $S$, $\bra Y_S \ket_t=\tr_S(\rho(t)Y_S)$ holds. 
Using $d\dke{\rho(t)}/dt=-iH\dke{\rho(t)} $ and \re{RI}, we obtain
\bea
\Dl E \ge \left \vert \f{\f{d}{dt} \sqrt{\bra X \ket_t} }{\sqrt{1-\bra X \ket_t}} \right \vert .
\eea
Then, 
\bea
L\defe \left \vert \int_0^\tau dt \ \f{\f{d}{dt} \sqrt{\bra X \ket_t} }{\sqrt{1-\bra X \ket_t}} \right \vert 
\le \int_0^\tau dt \ \left \vert \f{\f{d}{dt} \sqrt{\bra X \ket_t} }{\sqrt{1-\bra X \ket_t}} \right \vert 
\le \int_0^\tau dt  \ \Dl E \la{key}
\eea
holds. Here,
\bea
L=\vert \cos^{-1}  \sqrt{\bra X \ket}_\tau- \cos^{-1}  \sqrt{\bra X \ket}_0  \vert=\cos^{-1} \abs{\dbra \rho(\tau) \dke{\rho(0)}}.
\eea
Because $\abs{\dbra \rho(t) \dke{\rho(0)}} \le F(\rho(t),\rho(0))$ (Theorem 9.4 in Ref.\cite{NC}), 
we obtain the Mandelstam-Tamm relation 
\bea
\mL(\rho(\tau),\rho(0)) \le \cos^{-1} \abs{\dbra \rho(\tau) \dke{\rho(0)}} \le \int_0^\tau dt  \ \Dl E . \la{MT_re}
\eea

Next, we demonstrate \re{MT_type}. 
We utilize the following inequality ((9.110) in Ref.\cite{NC})
 \bea
 \half \nor{\rho-\sig}_1 \le \sqrt{1-[F(\rho, \sig)]^2}.
 \eea
 This leads to 
 \bea
 \half \nor{\rho-\sig}_1 \le \sin \mL(\rho, \sig) \le \mL(\rho, \sig).
 \eea
 This relation and \re{MT_re} lead to \re{MT_type}.

\section{Total entropy production rate and activities} \la{App_B}

\subsection{Semi-inner product}

If we diagonalize $\tl \rho(t)$ as 
\bea
 \tl \rho(t)\aeq \sum_n p_n(t)\ke{n(t)}\brv{n(t)} , \la{D_tl}
 \eea
we obtain
\bea
\dbra X, Y\dket_{\tl \rho, k}  \aeq \tr\Big[X\dg  \int_0^1 ds \  (\ga_{-k}\tl \rho)^s Y (\ga_k \tl \rho)^{1-s} \Big] \no\\
 \aeq \sum_{n,m} \brv{n}X\dg \ke{m} \brv{m}Y\ke{n} \int_0^1 ds \  (\ga_{-k}p_m)^s (\ga_k p_n)^{1-s} \no\\
 \aeq \sum_{n,m} M_k(m,n)  \brv{m}X\ke{n}^\ast \brv{m}Y\ke{n}. \la{semi}
\eea
Here, $M_k(m,n)\defe \Psi(\ga_{-k}p_m, \ga_{k}p_n) $ is a weight and $\Psi(a,b)$ is the logarithmic mean. 
Then, the semi-inner product corresponds to a weighted inner product introduced for the master equation in Ref.\cite{Ito2022}.
Note that $\nor{X}_{\tl \rho, k}^2 =0$ does not lead to $X=0$. 
Even if $\nor{X}_{\tl \rho, k}^2 =0$, $\brv{m}X\ke{n}$ can remain for $(m, n)$ such that $M_k(m,n)=0$.

\subsection{Total entropy production rate} \la{A_entropy}

In this subsection, we demonstrate that 
\bea
\dot{\sig}=\half \sum_k \Big\vert \Big\vert [\tl L_k,-\ln \tl \rho-\be_b \tl H_S] \Big\vert\Big\vert_{\tl \rho, k}^2.
\eea

As we will prove in the end of this section, 
\bea
\{\rho\}_k([X,\ln \rho]+\be_b \om X) \aeq \ga_kX\rho - \ga_{-k} \rho X  \la{L1}
\eea
holds. 
Using \re{L1}, we obtain
\bea
\ga_k\tl L_k \tl \rho -\ga_{-k} \tl \rho \tl L_k
\aeq \{\tl \rho\}_k([\tl L_k ,\ln \tl \rho]+\be_b \om \tl L_k)  \no\\
\aeq  \{\tl \rho\}_k([\tl L_k ,\ln \tl \rho+\be_b \tl H_S] ) .
\eea
Here, we used \re{Lind1} in the second line.
Using this, we obtain
\bea
&&\hs{-4mm}\half \sum_{a,\om} [\tl  L_k\dg, \{\tl \rho\}_k([\tl L_k ,-\ln \tl \rho-\be_b \tl H_S] )] \no\\
\aeq \half \sum_{a,\om} [\tl L_k\dg, -\ga_k\tl L_k \tl \rho +\ga_{-k} \tl \rho \tl L_k] \no\\
\aeq \half \sum_{a,\om} \Big(-\ga_k \{ \tl L_k\dg \tl L_k \tl \rho-\tl L_k \tl \rho \tl L_k\dg \} 
+\ga_{-k} \{\tl L_k\dg \tl \rho \tl L_k-\tl \rho \tl L_k \tl L_k\dg\} \Big)  \no\\
\aeq \half \sum_{a,\om} \ga_k\Big(- \tl L_k\dg \tl L_k \tl \rho+\tl L_k \tl \rho \tl L_k\dg 
+\tl L_k \tl \rho \tl L_k \dg-\tl \rho \tl L_k \dg \tl L_k\} \Big) \no\\
\aeq \tl \mD_b(\tl \rho). \la{eq_S25}
\eea
Here, we used \re{Lind1} in the third line, and $ \tl \mD_b(\tl \rho) \defe \sum_{a,\om}\ga_k \hat{D}[\tl L_k](\tl \rho)$.
\re{eq_S25} demonstrates that the Laplacian $\hat{O}^{(b)}(X) =\half \sum_{a,\om}[\tl L_k\dg,\{\tl \rho \}_k([\tl L_k,X]) ]$
acting on the operator of thermodynamic force ($X=-\ln \tl \rho-\be_b \tl H_S$) becomes the dissipator \cite{Ito2022}. 
Eventually, we relate the entropy production rate with the norms of the commutators:
\bea
\dot{\sig} \aeq  \sum_b \tr\{ ( -\ln  \rho-\be_b  H_S) \mD_b(\rho) \} \no\\
\aeq \sum_b \tr\{ ( -\ln \tl \rho-\be_b \tl H_S)\tl \mD_b(\tl \rho) \} \no\\
\aeq \sum_b \tr \Big[( -\ln \tl \rho-\be_b \tl H_S) \half \sum_{a,\om} [\tl  L_k\dg, \{\tl \rho\}_k([\tl L_k ,-\ln \tl \rho-\be_b \tl H_S] )]  \Big] \no\\
\aeq \half \sum_k \Big\vert \Big\vert [\tl L_k,-\ln \tl \rho-\be_b \tl H_S] \Big\vert\Big\vert_{\tl \rho, k}^2.
\eea

The equality \re{L1} is derived as follows:
\bea
&&\hs{-4mm}\{\rho\}_k([X,\ln \rho]+\be_b \om X) \no\\
\aeq \int_0^1 ds \  (\ga_{-k}\rho)^s (X \ln \rho -\ln \rho X-\be_b \om X)  (\ga_k \rho)^{1-s} \no\\
\aeq - \ga_k\int_0^1 ds \ \f{d}{ds}[e^{-s\be_b \om}e^{s\ln \rho} X e^{(1-s)\ln \rho}] \no\\
\aeq  \ga_k X\rho- \ga_{-k}\rho X  . \la{key_Vu}
\eea
Here, we used \re{Lind2}. 
\re{key_Vu} corresponds to (S5e) of Ref.\cite{Vu21}.

\subsection{Activity} \la{A_B}

In this subsection, we demonstrate that 
\bea
\mB(t) \defe \half \sum_k \Big\vert \Big\vert [\tl L_k,\varphi] \Big\vert\Big\vert_{\tl \rho, k}^2 \le 2A_\varphi(t).
\eea
Using \re{semi} and $\Psi(a,b) \le \f{a+b}{2}$, we obtain \cite{Vu21, Vu22}
\bea
 \nor{X}_{\tl \rho, k}^2 \aeqle \half \sum_{n,m} \Big(\ga_{-k}p_m\brv{n}X\dg \ke{m} \brv{m}X\ke{n} 
+\ga_{k}p_n\brv{n}X\dg \ke{m} \brv{m}X\ke{n}\Big) \no\\
 \aeq \half \Big[\ga_{-k}\tr\Big(\tl \rho XX\dg \Big)+ \ga_{k}\tr\Big(\tl \rho X\dg X\Big) \Big] .
\eea
Then, for $X_k \defe [\tl L_k,\varphi]$, 
\bea
\mB(t)  \aeqle \f{1}{4} \sum_k\Big[\ga_k \tr(\tl \rho X_k\dg X_k)
+ \ga_{-k}\tr(\tl \rho X_k X_k \dg ) \Big]  \no\\
\aeq \half \sum_k \ga_k \tr(\tl \rho X_k\dg X_k ) \no\\
\aeq 2A_\varphi(t)\la{mB}
\eea
holds. 
Here, we used \re{Lind1} in the second line of \re{mB}. 
Further by using 
\bea
\tr\big( \tl \rho(t) \{\varphi , \tl L_k\}\dg \{\varphi , \tl L_k\} \big) \ge 0,
\eea
$\varphi(t)^2=1$, and $B_\infty(t) \ge B^\pr(t)$, we obtain \re{eq41}.

\subsection{Partial activity} \la{A_C}

Because of \re{d_T_int},
\bea
\sig \aeqge\f{d_V(\rho(\tau), \rho(0))^2}{\int_0^\tau dt \ 2A_\phi(t)} \la{Vu-Hasegawa_2}
\eea
can be derived in the same way as \re{key_2} and \res{A_B}. 
$d_V(\rho(\tau),\rho(0))$ is defined by
\bea
d_V(\rho(\tau),\rho(0)) \defe \sum_n \abs{p_n(\tau)-p_n(0)} \la{d_V_SM}
\eea
using the spectral decomposition 
\bea
 \rho(t)\aeq \sum_n p_n(t)\ke{n(t)}\brv{n(t)} \la{A_SD}
\eea
with differentiable eigenstates $\{\ke{n(t)}\}$. 
$A_\phi(t)$ is given by \re{A_varphi} replacing $\varphi$ with $\tl \phi$: 
\bea
A_\phi(t) \aeq \tr\Big(\rho(t) \f{1}{4}\sum_k \ga_k [ \phi ,  L_k]\dg [ \phi , L_k] \Big).
\eea
Here, 
\bea
\phi(t) \aeqd \sum_n c_n(t) \ke{n(t)}\brv{n(t)},
\eea
and $c_n(t)=\RM{sign}(p_n(t)-p_n(0))$ or $c_n(t)=\RM{sign}(p_n(\tau)-p_n(0))$. 
We obtain
\bea
A_\phi(t) \aeq \sum_{n,m} \f{1}{4}\{c_m(t)-c_n(t)\}^2p_n(t)\sum_k \ga_k \abs{\brv{m(t)}L_k \ke{n(t)}}^2 \no\\
\aeq \sum_{c_n(t)  \ne c_m(t) } p_n(t)\sum_k \ga_k \abs{\brv{m(t)}L_k \ke{n(t)}}^2\no\\
\aeqle \sum_{n \ne m}p_n(t)\sum_k \ga_k \abs{\brv{m(t)}L_k \ke{n(t)}}^2 = \mA(t).
\eea
Here, $\mA(t)$ is given in \re{mA}. 
$A_\phi(t)$ corresponds to the partial activity \cite{Vu21}. 
In the classical master equation limit, $A_\varphi(t)$ becomes $A_\phi(t)$ with $c_n(t)=\RM{sign}(p_n(t)-p_n(0))$.

\section{\rd{Supplement for \re{tr_distance_s}}} \la{s_A}

We derive a bound for the trace distance in the Schr\"{o}dinger picture. 
This bound does not refer to the virtual isolated system, 
as opposed to the first term of the right-hand side of \re{tr_distance_s} referencing it.

From the quantum master equation \re{QME} and the triangle inequality, we obtain \cite{E9, Funo19}
\bea
\nor{\rho(\tau)-\rho(0)}_1 \aeqle \int_0^\tau dt \ \Big\vert \Big\vert \f{d\rho}{dt} \Big\vert \Big\vert_1 \no\\
\aeqle \int_0^\tau dt \ \Big(\nor{-i[H_S,\rho]}_1+ \nor{\mD(\rho)}_1 \Big).
\eea
For simplicity, in this section, we suppose that the dimension of the Hilbert space of the system is finite.
The polar decomposition \cite{NC} of $\mD(\rho)$ is given by
$\mD(\rho) = V\dg \sqrt{[\mD(\rho)]\dg \mD(\rho)}$ $(V \dg V=1)$.
This decomposition leads to 
\bea
\nor{\mD(\rho)}_1 = \tr(V\mD(\rho)) .
\eea
By repeating similar calculations from \re{key_1}, we obtain
\bea
\tr(V\mD(\rho)) \le \sqrt{\dot{\sig}}\sqrt{ \half \sum_k \Big\vert \Big\vert [\tl L_k,V] \Big\vert\Big\vert_{\tl \rho, k}^2} 
\eea
with 
\bea
\half \sum_k \Big\vert \Big\vert [\tl L_k,V] \Big\vert\Big\vert_{\tl \rho, k}^2 \le  2A_V(t) 
\le B(t) + B_V(t) \le B(t)+B_\infty(t)
\eea
where
\bea
 A_V(t) \aeqd \tr \Big(\rho \sum_k \ga_k \f{1}{4} [V, L_k]\dg   [V, L_k] \Big) 
\eea
and $B_V(t) \defe \tr \big(V \rho V\dg \sum_k \ga_k  L_k\dg  L_k \big)  $. 
Thus, we obtain a bound for the trace distance in the Schr\"{o}dinger picture:
\bea
\nor{\rho(\tau)-\rho(0)}_1 \le \int_0^\tau dt \ \nor{-i[H_S,\rho]}_1+\sqrt{\sig} \sqrt{\int_0^\tau dt \ 2 A_V(t)} .\la{5.7}
\eea
Because $\nor{-i[H_S,\rho]}_1 \le 2\Dl E$ \cite{Funo19}, the first term of the right-hand side is bounded by the Mandelstam-Tamm type term. 
The second term of the right-hand side corresponds to Shiraishi \textit{et al}.'s relation \re{Shiraishi_relation}
and $c_2 + c_3$ in \re{Funo_relation}.
In the classical master equation limit, 
$V$ becomes a diagonal matrix of which components are $\RM{sgn}(\f{dp_n}{dt})=\pm 1$. 
In this case, the first term of the right-hand side of \re{5.7} vanishes and $A_V \le A_\RM{c}$.

\section{Distances $d_T$ and $d_V$} \la{A_D}

The distance $d_T$ is defined by 
\bea
d_T(\rho(\tau), \rho(0)) \defe \sum_n \abs{b_n -a_n}
\eea
where $\{a_n \}$ and  $\{b_n \}$ are increasing eigenvalues of $\rho(0)$ and $\rho(\tau)$. 
$d_V(\rho(\tau),\rho(0))$ is defined by \re{d_V_SM} and can  be rewritten as
\bea
d_V(\rho(\tau),\rho(0)) = \sum_n \abs{b_{\chi{(n)}} -a_n}.
\eea
Here, $\chi$ is a permutation. 
For any two increasing sequences $\{x_n \}$ and  $\{y_n \}$, we can demonstrate that
\bea
\sum_n \abs{y_{\sig{(n)}} -x_n} \ge  \sum_n \abs{y_n -x_n}  \la{A_D_key}
\eea
for an arbitrary permutation $\sig$. 
Then, we obtain 
\bea
d_V(\rho(\tau),\rho(0)) \ge d_T(\rho(\tau),\rho(0)) .
\eea
If the eigenvalues of $\rho(t)$ don't intersect, $d_V(\rho(\tau),\rho(0)) = d_T(\rho(\tau),\rho(0)) $ holds. 

We prove \re{A_D_key}. 
For $i<j$ with $\sig(i) > \sig(j)$, we can prove that
\bea
\abs{y_{\sig(i)}-x_i}+\abs{y_{\sig(j)}-x_j} \aeqge \abs{y_{\sig(j)}-x_i}+\abs{y_{\sig(i)}-x_j} .
\eea
The above equation leads to \re{A_D_key} 
because $\{y_{\sig{(n)}}\}$ becomes the increasing sequence $\{y_n \}$ by repeating this type of exchanging.

\section{Detailed calculation for \res{s_Qubit}} \la{A_Qubit}

From \re{QME_Qubit}, we otain $L_1=2\sig_+(\theta(t))$, $\ga_1=\f{1}{4}\al \ga\ep(t) n(\ep(t))$, $L_2=2\sig_-(\theta(t))$, 
and $\ga_2=\f{1}{4}\al \ga\ep(t) [ n(\ep(t))+1]$. 
Denoting $U\dg(t) \tau_i(\theta(t))U(t)$ by $\sum_k T_{ik}(t)\tau_k$, we obtain 
$(\bm{R}_1)_i=(\bm{R}_2)_i=T_{1i}$, $(\bm{I}_1)_i=-(\bm{I}_2)_i=T_{2i}$, and $(\bm{R}_1 \times \bm{I}_1)_i=T_{3i}$. 
The equations of motion of $T_{ik}$ are given by
\bea
\f{dT_{1j}}{dt} \aeq -\ep(t)T_{2j}-T_{3j} \f{d\theta(t)}{dt} ,\\
\f{dT_{2j}}{dt} \aeq \ep(t)T_{1j} ,\\
\f{dT_{3j}}{dt} \aeq T_{1j} \f{d\theta(t)}{dt}.
\eea
The equation of motion of the Bloch vector is given by
\bea
\f{dr_i}{dt} =-\al \ga \ep(t)\Big([2n(\ep(t))+1] \Big[r_i -\sum_k\half \{ T_{1k} T_{1i} +T_{2k} T_{2i} \}r_k \Big] 
+T_{3i} \Big). 
\eea
The activities $A_\varphi(t)$, $B(t)$, $B^\pr(t)$, and $B_\infty(t)$ are given by
\bea
A_\varphi(t) \aeq \f{\al \ga \ep(t)}{4}\Big([2n(\ep(t))+1] \Big[1+(\bm{\varphi}\cdot \bm{T}_3)^2 \Big]
+2(\bm{\varphi}\cdot \bm{T}_3)(\bm{\varphi}\cdot \bm{r}) \Big) ,\\
B(t) \aeq \f{\al \ga\ep(t)}{2} \Big[2n(\ep(t))+1+x_3 \Big], \\
B^\pr(t) \aeq \f{\al \ga\ep(t)}{2} \Big[2n(\ep(t))+1-x_3+ 2(\bm{\varphi}\cdot \bm{T}_3)(\bm{\varphi}\cdot \bm{r})\Big],
\eea
and $B_\infty(t)=\Ga(t)\defe \al \ga\ep(t)[2n(\ep(t))+1] $.
Here, $(\bm{T}_3)_i \defe T_{3i}$ and  $x_i\defe \sum_k T_{ik}r_k$.
The equations of motion of $x_i(t)$ are given by
\bea
\f{dx_1}{dt} \aeq -\ep(t)x_2- \f{d\theta(t)}{dt}x_3 -\half \Ga(t)x_1 ,\\
\f{dx_2}{dt} \aeq  \ep(t)x_1-\half \Ga(t)x_2 ,\\
\f{dx_3}{dt} \aeq  \f{d\theta(t)}{dt}x_1-\Ga(t)\Big[x_3+\f{1}{2n(\ep(t))+1} \Big].
\eea
The heat current is given by
\bea
J(t) \aeq -\f{\ep(t)}{2}\sum_k T_{3k}\f{dr_k}{dt} \no\\
\aeq \f{\ep(t)}{2}\al \ga \ep(t) \Big( [2n(\ep(t))+1]x_3+1\Big) .
\eea
The entropy production is calculated by \re{sig_2}.


\begin{thebibliography}{99}
\bibitem{local_detailed_balance}$W_{nm}$ is given by $W_{nm}=\sum_b W_{nm}^{(b)}$ where $W_{nm}^{(b)}$ is the contribution from the bath $b$. 
The local detailed balance condition is $W_{nm}^{(b)}e^{-\be_b E_m}=W_{mn}^{(b)}e^{-\be_b E_n}$ where $E_n$ is the energy of the state $n$ and $\be_b$ is the inverse temperature of the bath $b$.
\bibitem{c_i}$c_1\defe \int_0^\tau dt\ \nor{-i[H,\rho]}_1$, $c_2\defe \int_0^\tau dt\ \nor{\mD_\RM{nd}(\rho)}_1$, and $c_3\defe \int_0^\tau dt\ \nor{\mD_\RM{d}(\rho)}_1$ where $\mD_\RM{d}(\rho) \defe \sum \ke{n} \brv{n}\mD(\rho) \ke{n} \brv{n} $ and 
$\mD_\RM{nd}(\rho) \defe \sum_{n \ne m} \ke{n} \brv{n}\mD[\rho] \ke{m} \brv{m}$. 
Here, $\{\ke{n}\}$ is defined by \re{SD}. 
Because of $\nor{-i[H,\rho]}_1 \le 2\Dl E$ \cite{Funo19}, \re{c_1} holds.
\bibitem{Q_to_C}If  (i) the eigenstate $\ke{E_n}$ of  $H_S$ are time-independent, (ii) $H_S$ is non-degenerate, 
(iii) the system state can be represented as $\rho(t)=\sum_n p_n(t)\ke{E_n}\brv{E_n}$, the quantum master equation reduces to the classical master equation. 
In this case, $\mW_{nm}$ corresponds to $W_{nm}$.
\bibitem{comment1}Precisely, in Ref.\cite{Vu21}, $\sig_\RM{V1}$ is  $\f{d_T(\rho(\tau), \rho(0))^2}{\int_0^\tau dt \ 2B_\infty(t)}$ where
$B_\infty(t) = \sum_k \ga_k \nor{L_k}_\infty^2 \ge B(t) $. 
However, $B_\infty(t)$ can be replaced by $B(t)$ as is done in Ref.\cite{Vu22}.
\bibitem{d_V}$d_T$ can be replaced with $d_V$ defined by $d_V(\rho(\tau),\rho(0)) \defe \sum_n \abs{p_n(\tau)-p_n(0)}$ using \re{SD} with differentiable eigenstates $\{\ke{n(t)}\}$. 
$d_V(\rho(\tau),  \rho(0)) \ge d_T(\rho(\tau), \rho(0))$ holds (\ref{A_D}). 
In the classical master equation limit, $d_V(\rho(\tau), \rho(0))$ becomes $l$. 
However, if the eigenvalues of $\rho(t)$ intersect as functions of time, it is difficult to find differentiable eigenstates or corresponding labeling.
\bibitem{jump_c}From the first equation of \re{Lind1} $[L_k, H_S]=\om L_k$, $\{L_k \}_{\om \ne 0}$ are traceless. 
For $\{L_k \}_{\om = 0}$, if  $L_k$ satisfy $[L_k, H_S]=0$, $L_k^\pr \defe L_k +l_k$ ($l_k$ are c-number satisfying $l_{-k}=l_k^\ast$) also satisfy they. 
And $\ga_k \hat{D}[L_k^\pr](Y)+\ga_{-k}\hat{D}[L_{-k}^\pr](Y)=\ga_k \hat{D}[L_k](Y)+\ga_{-k}\hat{D}[L_{-k}](Y)$ holds.  
Then, we can assume that $\{L_k \}_{\om = 0}$ are also traceless.
\end{thebibliography}
\end{document}